\documentclass[10pt,conference]{IEEEtran}

\usepackage{amsmath, amssymb, bm, cite, epsfig, psfrag}
\usepackage{graphicx}
\usepackage{subcaption}
\usepackage[top    = 0.7in,
bottom = 0.95in,
left   = 0.7in,
right  = 0.7in]{geometry}
\usepackage{array}
\usepackage{textcomp} 
\usepackage{longtable}
\usepackage{supertabular,booktabs}
\usepackage{authblk}
\usepackage{multirow}
\usepackage[usenames,dvipsnames]{xcolor}
\usepackage{etoolbox}
\usepackage{pbox}
\usepackage{romannum}
\usepackage{enumerate}
\usepackage{colortbl}
\usepackage{url}
\usepackage[switch]{lineno} 
\usepackage{bm}
\usepackage{float}
\usepackage{lipsum}
\usepackage{fancyhdr}
\graphicspath{{figures/}}

\newcommand{\figref}[1]{Fig.~\ref{#1}}
\fancypagestyle{firststyle}{
	\fancyhf{}
	\fancyhead[L]{A. Chopra, A. Thornburg, O. Kanhere, S. S. Ghassemzadeh, M. Majumdar, and T. S. Rappaport, ``A Real-Time Millimeter Wave V2V Channel Sounder,''  2022 \textit{IEEE Wireless Communications and Networking Conference (WCNC)}, April 2022, pp. 1-6.
	}                           

}
\IEEEoverridecommandlockouts

\begin{document}
\title{A Real-Time Millimeter Wave V2V Channel Sounder}

\author[1]{Aditya Chopra}
\author[1]{Andrew Thornburg}
\author[2]{Ojas Kanhere}
\author[1]{\\Saeed S. Ghassemzadeh}
\author[1]{Milap Majmundar}
\author[2]{Theodore S. Rappaport}
\affil[1]{AT\&T Labs, \{ac116g, at911r, 
		sg2121, mm1805\}@att.com}
\affil[2]{NYU Tandon School of Engineering, \{ojask, tsr\}@nyu.edu}

\maketitle
	\thispagestyle{firststyle}
\begin{abstract}
Wireless communication in millimeter wave spectrum is poised to provide the latency and bandwidth needed for advanced use cases unfeasible at lower frequencies. Despite the market potential of vehicular communication networks, investigations into the millimeter wave vehicular channel are lacking. In this paper, we present a detailed overview of a novel 1 GHz wide, multi-antenna vehicle to vehicle directional channel sounding and measurement platform operating at 28 GHz. The channel sounder uses two 256-element phased arrays at the transmitter vehicle and four 64-element arrays at the receiver vehicle, with the receiver measuring 116 different directional beams in less than 1 millisecond. By measuring the full multi-beam channel impulse response at large bandwidths, our system provides unprecedented insight in instantaneous mobile vehicle to vehicle channels. The system also uses centimeter-level global position tracking and 360 degree video capture to provide additional contextual information for joint communication and sensing applications. An initial measurement campaign was conducted on highway and surface streets in Austin, Texas. We show example data that highlights the sensing capability of the system. Preliminary results from the measurement campaign show that bumper mounted mmWave arrays provide rich scattering in traffic as well a provide significant directional diversity aiding towards high reliability vehicular communication. Additionally, potential waveguide effects from high traffic in lanes can also extend the range of mmWave signals significantly.
\end{abstract}
    

\section{Introduction}\label{sec:Introduction}
Fifth generation (5G) and beyond networks are designed for many different use-cases and verticals. Future networks will enable billions of connections with varying geographic and mobility requirements \cite{Rappaport_2013b}. Vehicular communication is one of the key verticals and use-cases of 5G communication standards. Sidelink vehicle to vehicle (V2V) communication is a feature of the Third Generation Partnership Projects' (3GPP) Release 16 \cite{3GPPv2x}. Sidelink communication enables new communication frameworks in a vehicular setting, such as mobile broadband range extension via relaying, or sensor sharing over peer-to-peer links. Applications such as platooning, collision avoidance, and autonomous driving are all expected to greatly benefit from the high throughput and low latency communications unlocked using mmWave band sidelink \cite{Va_2016}. 

To realize these vehicular applications, the intricate behavior of the mmWave wireless channel in a vehicular communication setting must be known. The sidelink as a V2V channel will exhibit different properties than as a vehicle to infrastructure (V2I) channel. There is a lack of mmWave V2V measurements in literature, where communication is dependent on directional transmissions using active phased array antennas to overcome the blockage due to objects and to overcome the omni-directional path-loss \cite{rappaport_2015}. Measuring V2V channels additionally presents a difficult problem of estimating a rapidly time-varying channel in high-speed scenarios, requiring real-time measurement durations on the order of microseconds. Furthermore, depending on the deployed height of the mmWave antenna systems, the transmitter and receiver may be close to the ground while also surrounded by many different metallic and reflective car and street furniture surfaces. Indeed, the focus of this effort is to characterize the V2V bumper mounted channel. This paper addresses two lacking characteristics of previous efforts: arrays mounted on vehicle bumpers and rapid, high bandwidth channel impulse response (CIR) capture.

 Much of the previous state-of-the-art channel measurement work uses passive directional antennas. Sliding correlator channel sounders have been used to extensively model mmWave channels over the last decade \cite{MacCartney_2015a,MacCartney_2015b,Rappaport_2019,Ben_Dor_2011}. Mechanical gimbals are used to adjust the orientation of antennas in azimuth and elevation over the course of seconds. Because of the use of gimbals and passive directional antennas, however, it is difficult to capture the rapidly changing channel. The authors of \cite{Ben_Dor_2011} conducted vehicular angle of arrival measurements in a parking lot environment. The authors of \cite{Huang2020} built a MIMO mmWave channel sounder using RF switching to rapidly measure the channel at 28, 32, and 39 \rm{GHz}. In addition to static human and vehicular blockage scenarios, the dynamic V2V channel was also measured, with relative TX-RX velocities varying from 0 km/h to 60 km/h, however, horn antennas were used for the measurements due to which the authors could only observe reflected paths arriving from four directions \cite{Huang2020}. The V2I channel at 60 \rm{GHz} was characterized with open-ended waveguide antennas in \cite{Blumenstein_2018}. Typical power delay profiles were measured and an rms delay of 1.5 ns was observed in the presence of vehicles \cite{Blumenstein_2018}. The work of \cite{Bas_2019} uses an active phased array to completely measure the channel in 1.44 ms, but the channel measured is a pedestrian channel where cars are driven between the TX and RX. In \cite{Caudill_2021},  dual-polarized 8 $\times$ 8 phased array antenna at 28.5 \rm{GHz} was used to develop an omnidirectional channel sounder, with the RX having a field-of-view of 360\textdegree\ in the azimuth and $ \pm $25\textdegree\ in the elevation. Channel measurements, however, were only conducted in a laboratory \cite{Caudill_2021}. Unlike previous channel sounders, the omnidirectional mmWave phased array channel sounder at 28 \rm{GHz} we present in this paper was used to collect real time (1 ms to complete an omnidirectional channel measurement) wide bandwidth measurements of the V2V channel in highly dynamic environments, with relative motion between the transmitter and receiver.

The remainder of this paper is organized as follows. Section \ref{sec:sys_design} describes the design of channel sounder transmitter and receiver subsystems. The measurement campaigns are described in Section \ref{sec:measure}. Initial results are presented in Section \ref{sec:results} followed by conclusions and future work in Section \ref{sec:conclusion}.

\section{System Design}\label{sec:sys_design}

\begin{figure}
\includegraphics[width=0.45\textwidth]{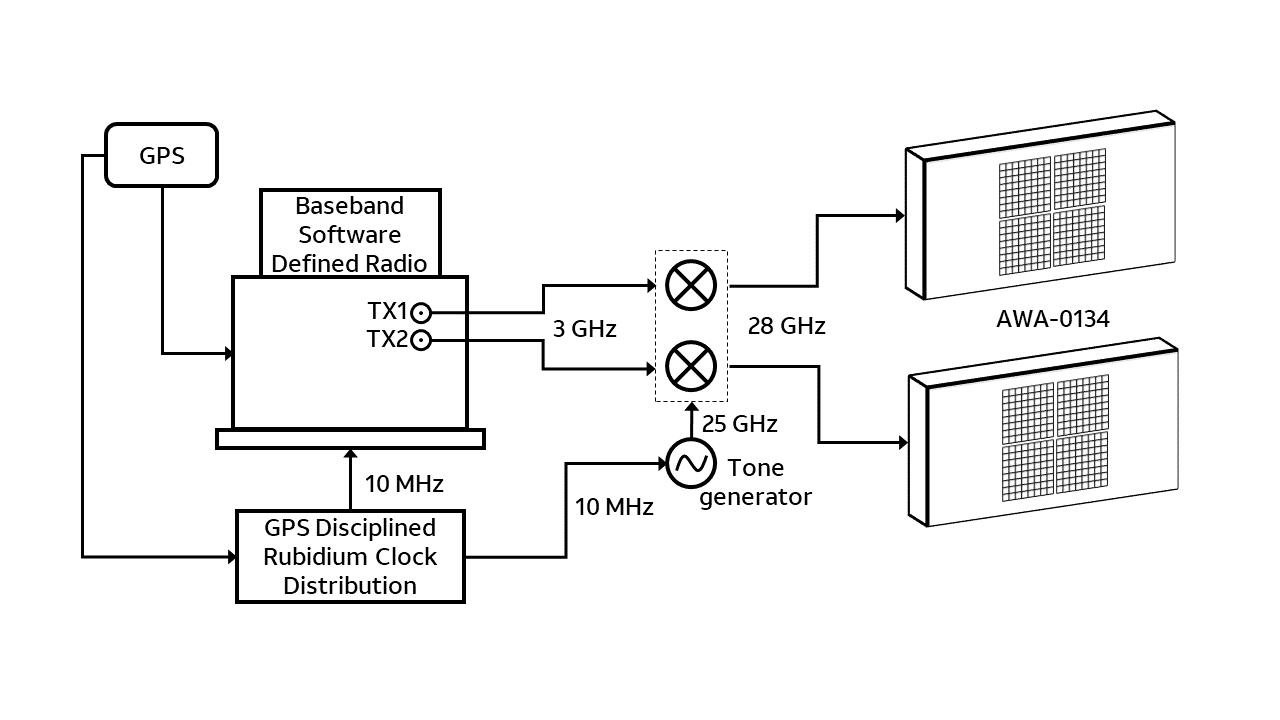}
\centering
\caption{Functional System diagram of the vehicular testbed two channel transmitter.}
\label{fig:tx_sys}
\end{figure}

\begin{figure}
  \includegraphics[width=0.475\textwidth]{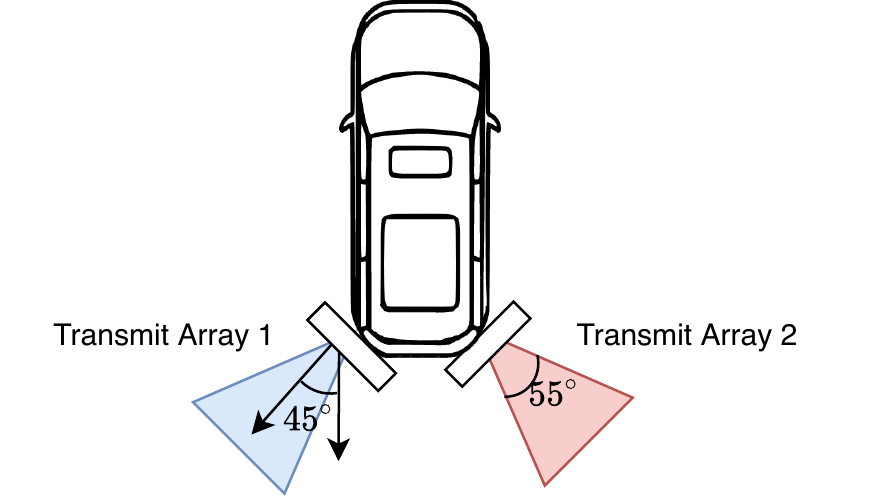}
  \centering
  \caption{Placement of transmit phased antenna array modules on rear bumper of van at a height of 15 inches.}
  \label{fig:tx_van}
  \end{figure}

The AT\&T Real-time Omni-directional Array Channel Sounder (ROACH) was originally described in \cite{Chopra_2020} \cite{kanhere_2021}. ROACH comprised of a single transmitter and a vehicle roof mounted receive subsystem with four phased arrays, that could measure the 28 GHz wireless channel over 60 MHz of bandwidth. In this paper, we present the next-generation iteration of the platform which greatly exceeds the capabilities of the previous version by utilizing wideband software defined radios (SDRs) and high-speed data recorders for real-time capture of 1~GHz wide channel impulse response measurements. Additionally, it can simultaneously capture channel measurements between two transmit and four receive antennas, resulting in $2\times4$ MIMO channel measurements. 

We retain the general channel sounding philosophy from our previous ROACH design by utilizing Zadoff-Chu (ZC) sequences~\cite{silva_2018} as sounding waveforms. ROACH transmits the ZC sequence $x_u[n]$ defined as
\begin{align}
  &x_u[n]=\text{exp}\left(-j\frac{\pi un(n+1)}{N_\text{ZC}}\right), \\
  \text{where}~ &0 \le n < N_\text{ZC},  \nonumber \\ 
  &0 < u < N_\text{ZC}~\text{and}~\text{gcd}(N_\text{ZC},u)=1.  \nonumber
\end{align}

\begin{figure}
\includegraphics[width=0.475\textwidth]{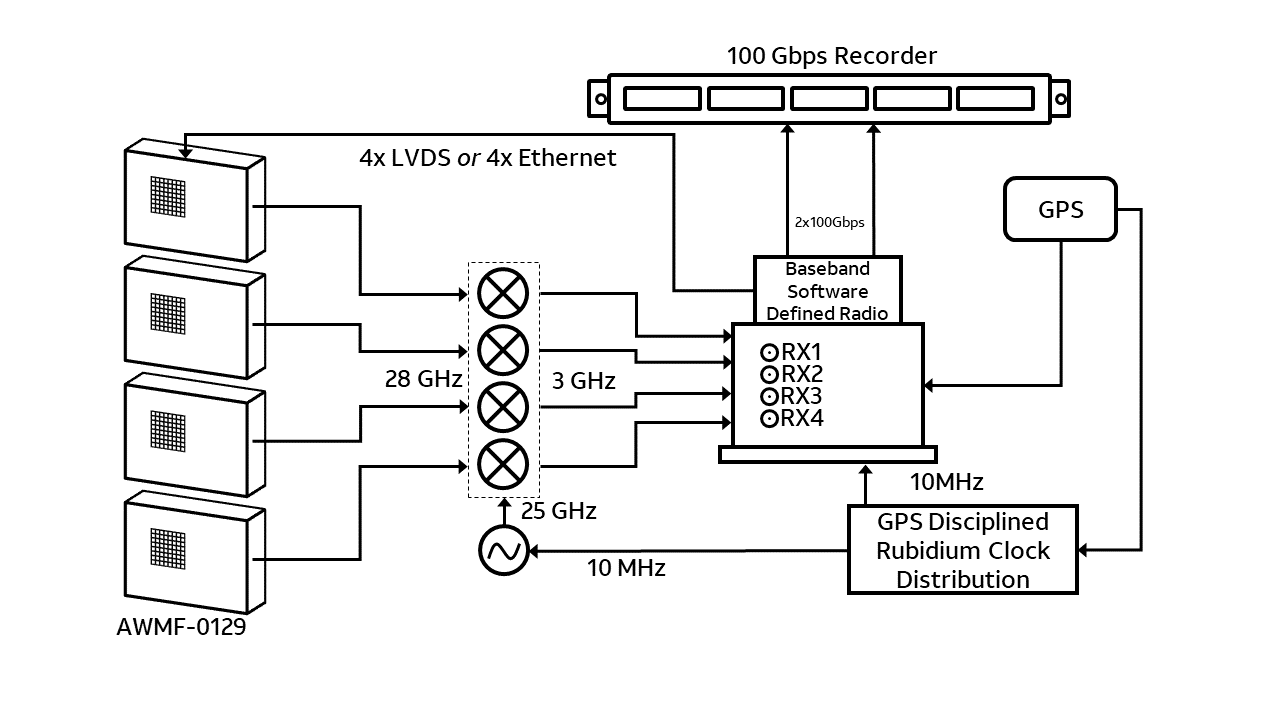}
\centering
\caption{Functional system diagram of vehicular testbed four channel receiver.}
\label{fig:rx_sys}
\end{figure}

\begin{figure}
  \includegraphics[width=0.475\textwidth]{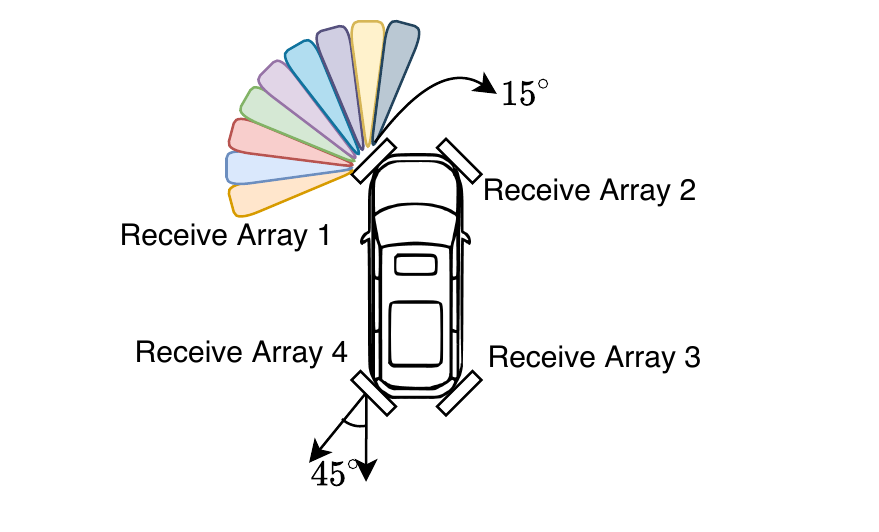}
  \centering
  \caption{Diagram showing placement of RX phased antenna array modules on van at a height of 15 inches.}
  \label{fig:rx_van}
  \end{figure}

  ZC sequences have constant amplitude and zero auto-correlation (CAZAC) and are hence commonly used in channel sounding, where low autocorrelation provides high signal to noise ratio (SNR) estimates of the CIR after wideband correlation, and constant amplitude allows the RF hardware to operate at or near power amplifier saturation, which increases overall efficiency and dynamic range of the sounder \cite{hua_2014}. The zero auto-correlation properties also allow our system to send multiple cyclically shifted orthogonal sequences for multi-antenna sounding. 
  
\begin{figure*}[ht]
  \includegraphics[width=0.75\textwidth]{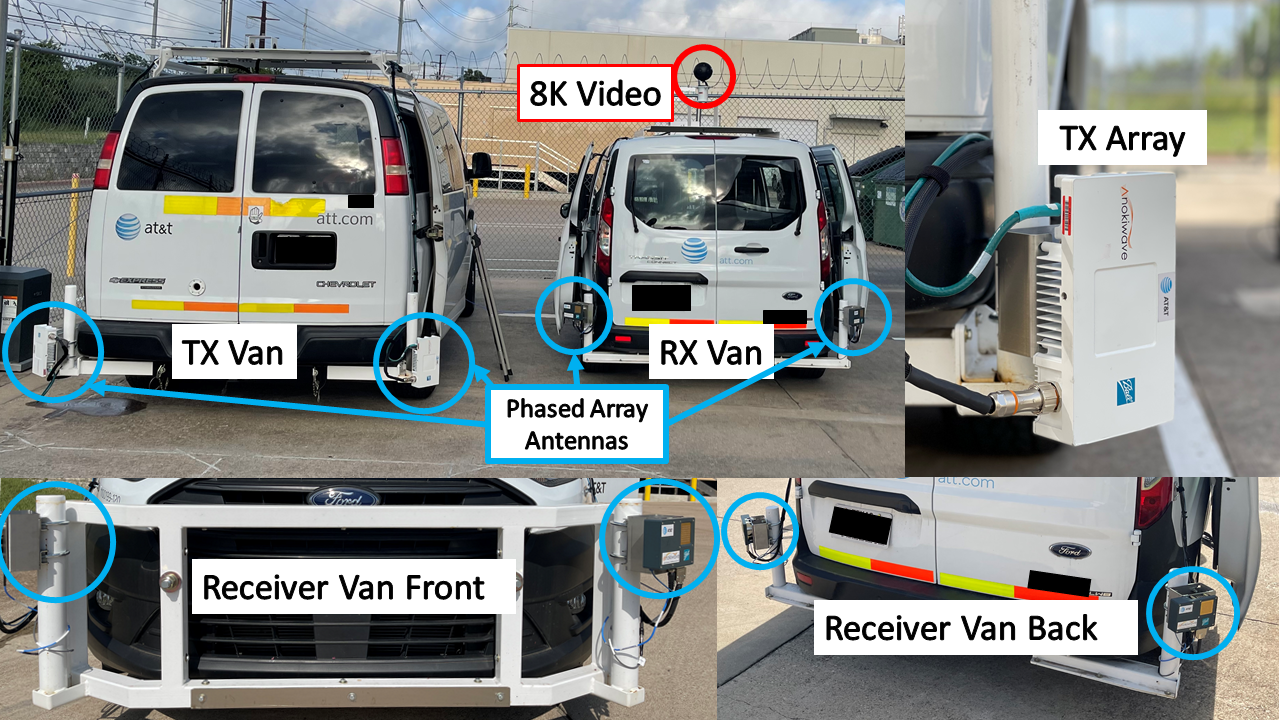}
  \centering
  \caption{The arrays are mounted at bumper level. The 64-element RX van has four arrays, one at each corner of the vehicle. The TX van has only two 256-element arrays, at the back of the vehicle.}
  \label{fig:vehicles}
  \end{figure*}

\subsection{Transmitter Architecture}

The overall transmitter architecture of ROACH is presented in Fig.~\ref{fig:tx_sys}. We utilize a Pentek 6350 baseband SDR~\cite{pentek}, which has eight analog to digital converters (ADCs) sampling at 4 gigasamples per second (GSps) as well as eight digital to analog converters (DACs) sampling at 6.5 GSps. These converters are directly connected to a Xilinx Ultrascale+ field programmable gate array (FPGA) compute fabric for waveform storage and processing. A benefit of using a flexible SDR is that while the results in this paper utilize ZC sequences, the ROACH system can load and transmit arbitrary waveforms of arbitrary length. This allows us to also use ROACH to emulate communication via standards based waveforms to prototype and test vehicular communication applications.

The baseband channel sounding waveform is internally upsampled before digital to analog conversion using the 6.5 GSps DAC to generate a direct conversion intermediate frequency (IF) signal with a bandwidth of 1 GHz. Using a numerically controlled oscillator (NCO), the IF signal can be placed in either the first or second Nyquist zone between 1 to 4 GHz center frequency. The subsequent IF signal is filtered by a band-pass analog filter to remove the harmonics in the other Nyquist zones.

The ROACH system utilizes two transmit channels in order to better understand spatial characteristics of the wireless V2V channel. Each of the transmit channels emits a complex ZC sequence of length 2048 and a symbol width of 1 ns, cyclically shifted to ensure orthogonality due to the zero autocorrelation properties of ZC sequences.

The two IF signals generated from the SDR module are upconverted to 28 GHz and transmitted by two Anokiwave AWA-0134 phased array systems as shown in Fig.~\ref{fig:tx_van}. Each of these is a 256 element active phased array antenna with a maximum overall gain of 59.1 dB~\cite{Chopra_2020}. At the transmitter, the Pentek SDR and both upconversion stages are driven from the same GPS disciplined 10 MHz reference. The sounding reference signal at the transmitter is generated every second using the pulse per second (PPS) signal from GPS satellites.

The two transmit arrays were mounted on the rear bumper of a van. The TX arrays were set to a static beam pointed outwards from each corner of the bumper, as shown in Fig.~\ref{fig:tx_van} and Fig.~\ref{fig:vehicles}. The 3~dB beamwidth of this beam was 55$^\circ$. Low loss mmWave cabling is used to minimize the loss from the bumper to the baseband. Power and control signaling for the phased arrays is routed from the bumper to a central rack within the van. The parallel mmWave signals fed into each of the phased arrays are calibrated to a power level of \text{-30~dBm}, resulting in an effective isotropic radiated power (EIRP) slightly lower than 30~dBm out of the antenna arrays.



\subsection{Receiver Architecture}

\begin{figure*}
\includegraphics[width=\textwidth]{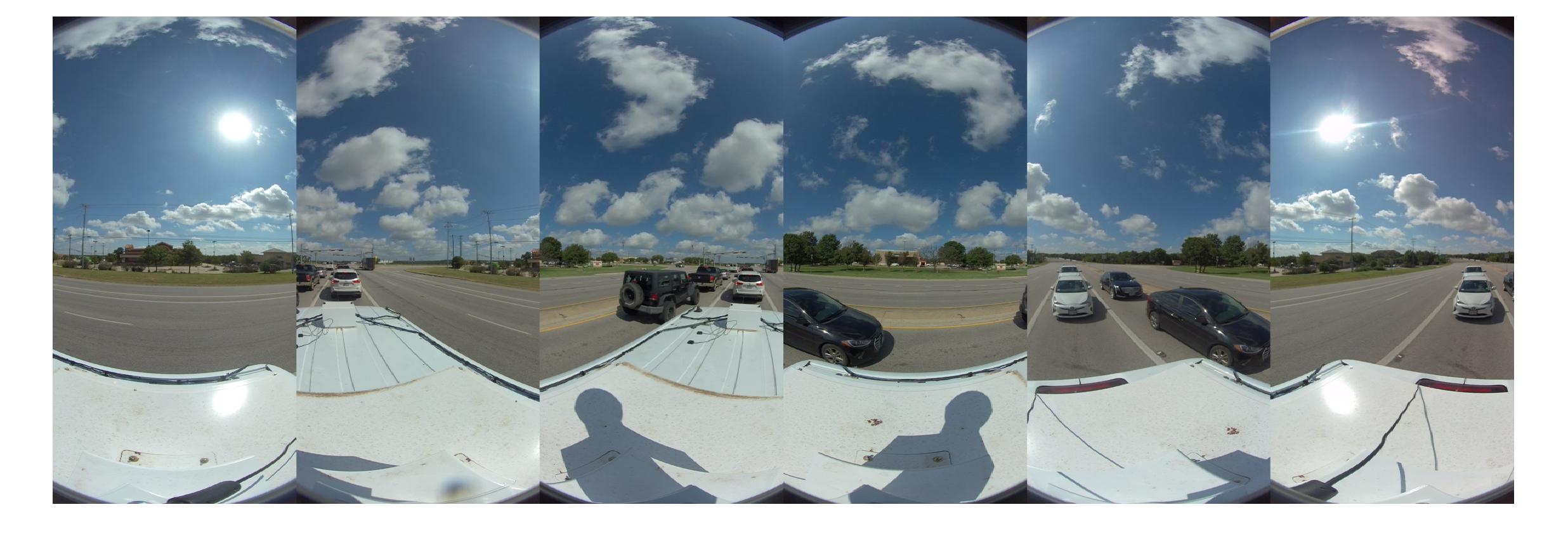}
\centering
\caption{An example video frame from the 360$^\circ$ camera. }
\label{fig:frame}
\end{figure*}

The ROACH testbed receiver comprises of an identical Pentek SDR capturing an IF signal downconverted from four mmWave receive channels at a (complex) sampling rate of 1 Gsps. The signals at each are captured and timestamped by the SDR, the resulting data is encapsulated into Ethernet packets. These packets are sent over 100GBE optical transceivers via UDP/IP interface directly to a 100 Gbps packet recorder. The data recorder operates at burst rates of 200 Gbps as well as sustained rates of 100 Gbps. GPS-aligned fractional seconds are stored with every CIR capture (each CIR being $ \approx $ 2 $ \mu s $ long) to correlate channel measurements with sensor data. 

Four Anokiwave AWMF-0129 phased array antenna modules are mounted at bumper height at each corner of the receive vehicle, as shown in Fig \ref{fig:rx_van}. All arrays have 64 elements with a maximum overall array gain of 47 dB~\cite{Chopra_2020}. The phased arrays were mounted at bumper height in contrast to the roof mounted arrays in the previous version of ROACH~\cite{Chopra_2020} since car designers are looking to improve aerodynamic efficiency, due to which additional bumps on the roof are undesirable. Channel measurements with the transmitter and receiver at vehicle bumper height suffer from more scattering, reflection, and blocking compared to roof mounted measurements. 

Each phased array sweeps at high speed through a beam codebook covering its quadrant around the vehicle. The four arrays simultaneously sweep through a codebook of 29 beams, dwelling $40~\mu s$ on each beam resulting in a 116 beam covering 360$^\circ$ of channel measurement in 1.16 ms. The codebook consisted of 29 equispaced beams per array covering $\pm 30^\circ$ in elevation and $\pm 45^\circ$ in azimuth; therefore all four arrays together grant $360^\circ$ azimuthal view of the V2V channel. This rapid beam switching speed in channel measurement is key to understanding the fast changing mmWave channel in a high mobility time-varying vehicular environment. Fig.~\ref{fig:rx_van} and Fig.~\ref{fig:vehicles} show the placement of the arrays and their coverage quadrants.

\subsection{Additional Sensors}
The ROACH system uses GPS modules at both the transmit and receive vehicles to accurately log the location and velocity of the vans as they drive. Using real time kinematic (RTK) corrections, the position of the vans is centimeter-level accurate, under typical situations. The position update rate for the GPS modules is 14 Hz (roughly every 70 ms).GPS is also used to maintain time and frequency synchronization among the multiple transmit and receive channels, as well as between the transmitter and receiver. 

On top of the receiver vehicle, we mounted an Insta360~Pro~2, 360$^\circ$ 8K camera. This camera module also uses GPS to time tag video frames, effectively synchronizing the video to the channel measurement data captured by our baseband system. These video frames can be used to correlate channel measurements with blocking and reflecting objects in the environment. Fig.~\ref{fig:frame} shows an example video frame from the first measurement campaign. This video frame is also used in the results in Section~\ref{sec:results} to show some of the capabilities of the ROACH system. 

\section{Vehicular Channel Measurement campaign}\label{sec:measure}

\begin{figure}[hb]
  \includegraphics[width=0.475\textwidth]{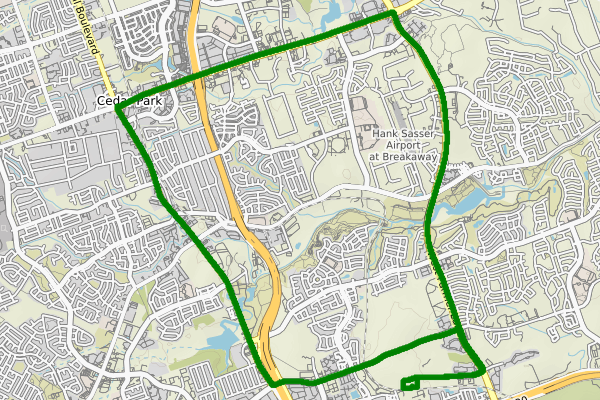}
  \centering
  \caption{Map of the RX van location traveling over a distance of 14.3 miles, at an average speed of 20 mph, and a peak speed of 60 mph on the first day.}
  \label{fig:drive}
  \end{figure}

We conducted an outdoor vehicular channel measurement campaign in Austin, Texas in the 28 GHz spectrum under an experimental license granted by the Federal Communication Commission~\cite{STA}. Fig.~\ref{fig:drive} shows an example subset of the drive map on one of the measurement days. The measurement campaign was conducted over multiple days and included both highway, urban, and suburban surface street drive tests. Each of the measurement drives contained a mixture of light and heavy traffic, slow and fast moving vehicles around the transmitter and receiver, as well as fixed and mobile reflectors in the environment. The synchronized video recording capabilities of ROACH are extremely useful in providing such environmental context to the channel measurements. The overall rate of data capture including channel measurements, video, and location data comes out to approximately one and a half terabytes per hour of drive time.

\begin{figure}
  \centering

\subfloat[Distance between transmit and receive vehicle.]{%
  \includegraphics{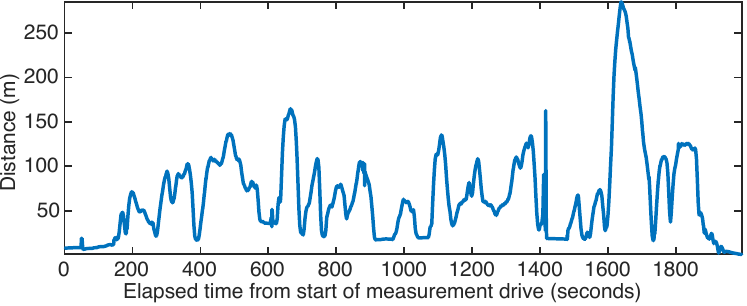}%
  \label{fig:tr-dist}%
}
\par\bigskip\medskip
\subfloat[Speed of the receive vehicle.]{%
  \includegraphics{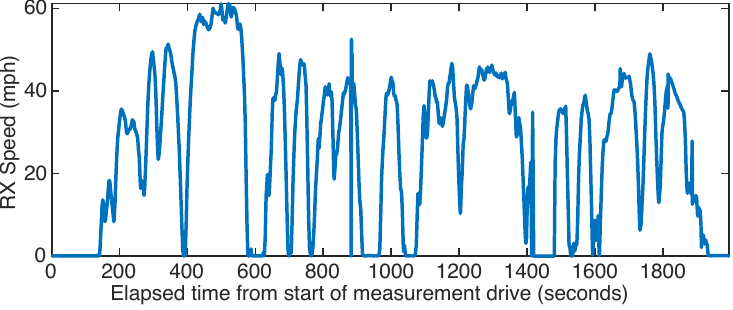}%
  \label{fig:rx-speed}%
}
\par\bigskip\medskip
\subfloat[Relative speed difference between transmit and receive vehicle.]{%
  \includegraphics{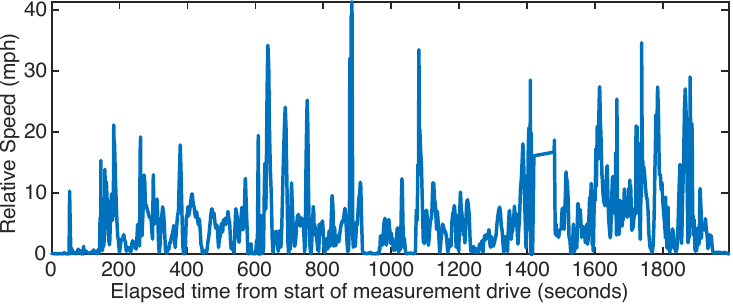}%
  \label{fig:rel-speed}%
}
\caption{Plot of various transmit and receive vehicle parameters over the duration of a single measurement drive run.}
\label{fig:stats}
\end{figure}

Fig. \ref{fig:stats} shows the key measurement drive characteristics from the example drive mapped in Fig.~\ref{fig:drive}. We can see a large range of distances between the transmitter and receiver, going as high as 250 meters in Fig.~\ref{fig:tr-dist}. The velocity map in Fig.~\ref{fig:rx-speed} also shows that both highway and surface street level of speeds were covered in the drive. A wide range of relative speeds observed in Fig.~\ref{fig:rel-speed} indicates a variety of Doppler conditions are captured in our channel measurements.




\section{Initial Results} \label{sec:results}

\begin{figure}[ht]
  \includegraphics[width=0.475\textwidth]{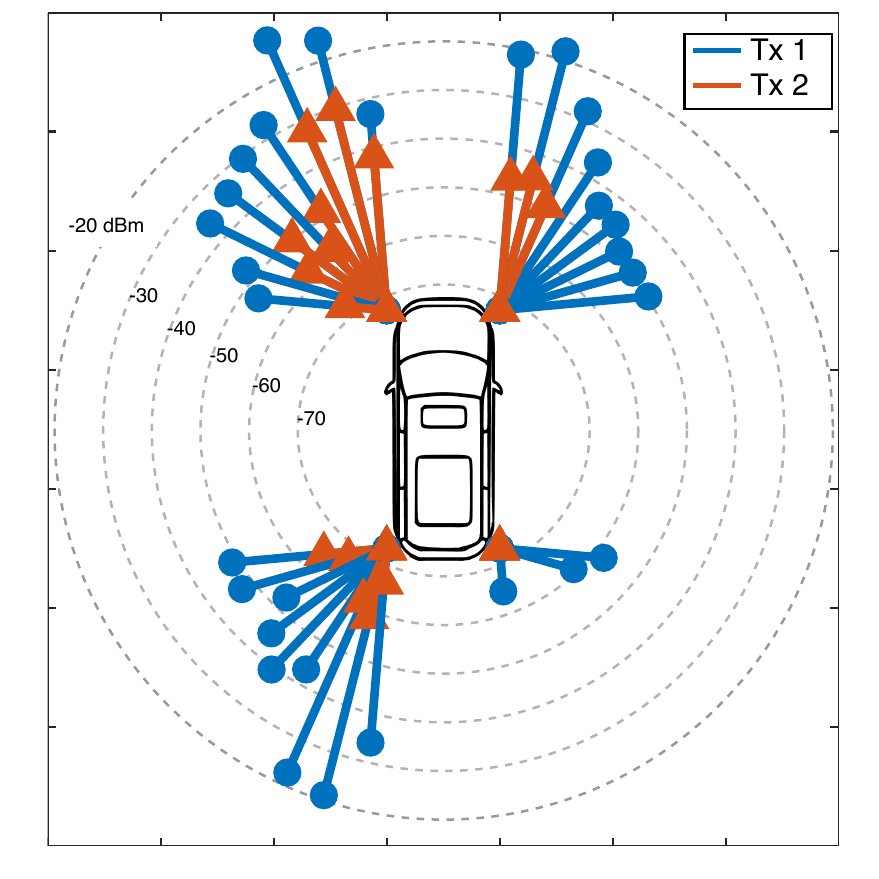}
  \centering
  \caption{Received signal strength in dBm between two transmitters and four receivers measured over different receive beams.}
  \label{fig:beam-360}
  \end{figure}

In this section, we present some initial findings from the first measurement campaign with the new ROACH system. We focus on the time snapshot shown in the video frame capture in \figref{fig:frame}. All preliminary results shown in this paper are obtained from a beam sweep at this time snapshot. Recall that at this point in time, the transmitter and receiver are blocked by other vehicles in between them, and there are vehicles on the left side of them and no vehicles on the right side. 
\figref{fig:beam-360} shows the power received from the different transmitters on all four receiver phased arrays. For sake of readability, we only consider beams from our codebook swept in azimuth while keeping the elevation fixed at 0$^\circ$. High amounts of coupling between transmitter 1 and the receivers placed on the left side of the vehicle is visible, indicate the presence of strong reflections from other traffic, even in a blocked direct path scenario. This theory is further bolstered by the fact that the coupling with transmitter 2 as well as coupling on the right and rear side receivers is much lower. 

\begin{figure}[ht]
    \includegraphics[width=0.45\textwidth]{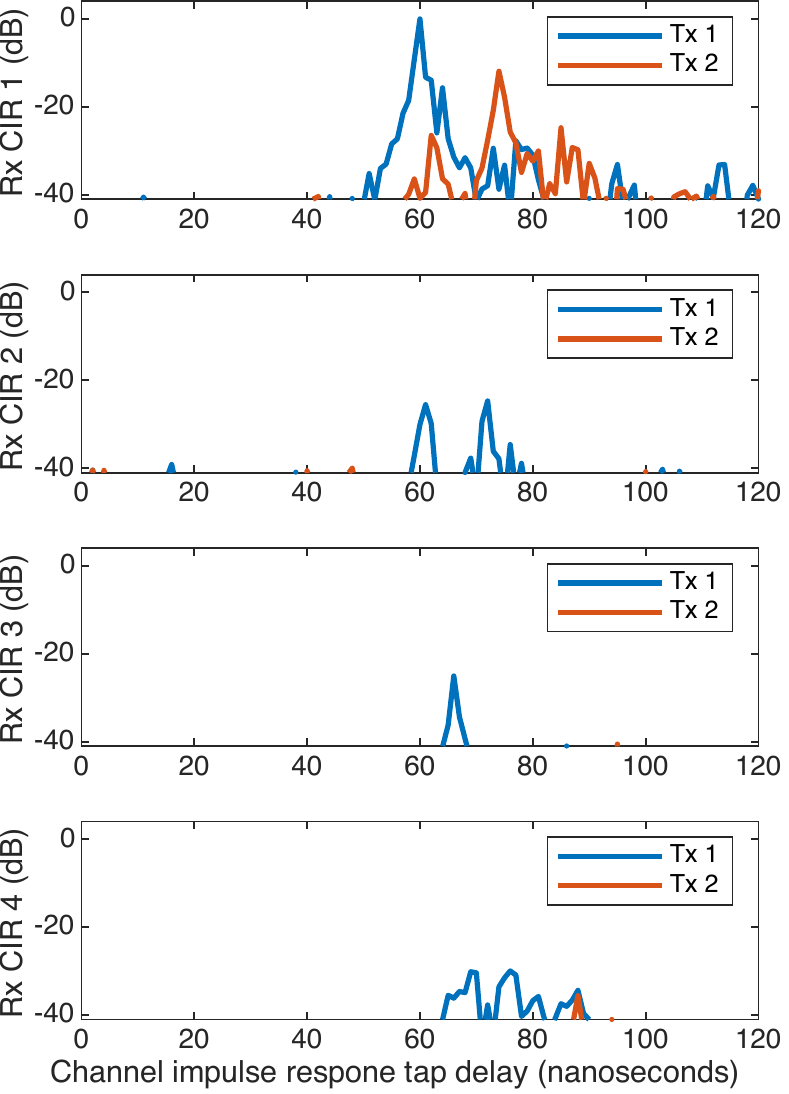}
    \centering
    \caption{Wireless CIRs from two transmitters and four receivers in a V2V communication scenario. The CIRs are normalized to the peak tap value which in this case is seen between transmitter 1 and receiver 1.}
    \label{fig:cir-stacked}
    \end{figure}

We further investigate the best beam on each of the four receivers and plot the strength of the channel measured from each of the two transmitters in \figref{fig:cir-stacked}. The absolute power of the CIR taps is plotted in dB and is normalized to the peak CIR tap value which in this case is observed by receiver 1. In the front side left array (receiver 1), one can see the direct path from both transmitters as well as multiple reflected paths. It is interesting to note that the reflections in case of transmitter 2 are around 20 dB stronger than the main path. In the rear of the receive vehicle, the left side array (receiver array 3) shows large smearing of reflected paths from transmitter 1 in its CIR as well. Keen eyed observers will notice a small time delay in receiver 3 CIR, which can be attributed to the 15 feet of vehicle length separation between the front arrays and the back arrays. It is interesting to note that only receiver 1 is able to capture signal from transmitter 2, and the signal reflected from other vehicles is stronger than the signal in the direct blocked path. This suggests a possible waveguide effect caused by multiple vehicles lined up next to each other with a small gap in between. This snapshot indicates that congested traffic conditions may help improve vehicular communication by providing a rich collection of reflective surfaces for the wireless signal.

\section{Conclusions and Future Work}\label{sec:conclusion}

In this paper we introduced the second-generation of ROACH, a one of a kind wideband omnidirectional mmWave vehicular channel sounder and testbed. The ROACH receiver can capture and record directional CIRs at a RF bandwidth of 1 GHz, covering 116 beams in under 1.2 milliseconds. ROACH also records location, speed, and environmental information that is synchronized with CIR measurements.  The measurements taken with ROACH during drive test campaigns in Austin, Texas present new insights into the V2V channel from a vehicle bumper height perspective. Auxiliary location and video data allows the channel measurements to be correlated with environmental scatters, blockers, and reflectors. We described the design of the overall channel sounding system, showed the capabilities of the system to do real-time recording, and provided preliminary analysis of directional V2V mmWave channel in a reflective environment highlighting the sensing capabilities of the wideband mmWave channel. 

In the future, we will present more insights into the V2V channel statistics and beam management in order to facilitate next generation vehicular applications and use-cases. Machine learning algorithms that can estimate traffic density from video are currently in development, and can provide key insight into the correlation of channel statistics to traffic levels. It is also important to study the impact of ground bounce or reflections on the wireless channel when the transmitter and receiver have one or more blocking vehicles in the middle.  Another avenue of future research is conducting drive campaigns during rain in order to provide novel insights into the behavior of the mmWave vehicular channel during adverse weather conditions.  

\bibliographystyle{IEEEtran}
\bibliography{references}{}

\end{document}